\documentclass[%
 reprint,
superscriptaddress,
 amsmath,amssymb,
 aps,
]{revtex4-1}

\usepackage{tabularx}
\usepackage{booktabs}
\usepackage{upgreek}
\usepackage{nicefrac}
\usepackage{hyperref}
\usepackage{natbib}
\usepackage{xcolor}
\usepackage{graphicx}
\usepackage{dcolumn}
\usepackage{bm}

\begin{document}

\preprint{APS/123-QED}

\title{Kinetic-beam-energy determination via collinear laser spectroscopy}

\author{Kristian K\"onig}
 \email{koenig@frib.msu.edu}
\affiliation{National Superconducting Cyclotron Laboratory, Michigan State University, East Lansing, USA}

\author{Kei Minamisono}
\affiliation{National Superconducting Cyclotron Laboratory, Michigan State University, East Lansing, USA}

\author{Jeremy Lantis}
\affiliation{National Superconducting Cyclotron Laboratory, Michigan State University, East Lansing, USA}

\author{Skyy Pineda}
\affiliation{National Superconducting Cyclotron Laboratory, Michigan State University, East Lansing, USA}

\author{Robert Powel}
\affiliation{National Superconducting Cyclotron Laboratory, Michigan State University, East Lansing, USA}

\date{\today}

\begin{abstract}
An approach to determine the kinetic beam energy at the $10^{-5}$ level is presented, which corresponds to an improvement by more than one order of magnitude compared to conventional methods. Particularly, collinear fluorescence and resonance ionization spectroscopy measurements on rare isotope beams, where the beam energy is a major contribution to the uncertainty, can benefit from this method. The approach is based on collinear spectroscopy and requires no special equipment besides a wavelength meter, which is commonly available.
Its advent is demonstrated in a proof-of-principle experiment on a Ni beam. In preparation for the energy measurement, the rest-frame transition frequencies of the $3d^9 4s\;^3\mathrm{D}_3 \rightarrow 3d^9 4p\;^3\mathrm{P}_2$ transitions in neutral nickel isotopes have been identified to be $\nu_0(^{58}\mathrm{Ni})=850\,343\,678\,(20)$\,MHz and $\nu_0(^{60}\mathrm{Ni})=850\,344\,183\,(20)$\,MHz.

\end{abstract}

\maketitle

\section{Introduction}
\label{sec:Introduction}

Collinear fluorescence and resonance-ionization laser spectroscopy (CLS) are well-established techniques for the measurement of molecular, atomic, and nuclear observables \cite{Otten.1989, Campbell.2016, Neugart.2017, Garcia.2020}. With widely tunable lasers, optical transitions in stable and short-lived isotopes can be accessed in flight at beam energies of typically $10-60$\,keV. 
The acceleration to this beam energy leads to a strong compression of the velocity width and enables Doppler-broadening-free measurements with a resonance-peak width at the level of the natural linewidth \cite{Kaufman.1976}. Furthermore, the fast beam velocity and in-flight detection make this technique, almost exclusively, the method of choice to access short-lived exotic isotopes. 
Measurements of hyperfine spectra of these isotopes are of high interest due to the direct link to the nuclear charge radius and electromagnetic moments, leading to investigations of, e.g., the nuclear shell structure \cite{Otten.1989, Campbell.2016}, the nuclear superfluidity \cite{Miller.2019}, the odd-even staggering in charge radii \cite{deGroote.2020}, the halo nuclei \cite{Geithner.2008, Noertershaueser.2009, Krieger.2012}, and the nuclear equation of state \cite{Yang.2018,Brown.2020}.

Due to the collinear geometry, the fast atoms experience a Doppler-shifted laser frequency. Hence, it is of critical importance to accurately determine the kinetic beam energy to correctly transform the observed resonance spectra into the rest frame of the atomic beam, where the nuclear information can be extracted from the isotope shifts and from the hyperfine splitting. Even though the impact of systematic uncertainties largely cancels in these relative measurements, the beam-energy uncertainty remains a dominant contribution. 
Under typical conditions, it is of the same order of magnitude as the statistical uncertainty that is limited by low production rates of radioactive isotopes (a few 10’s ion/s for the rarest isotopes that were investigated with CLS so far \cite{Miller.2019, deGroote.2020}).
The beam energy is usually determined by a direct measurement of the acceleration potential or indirectly by measuring an isotope shift that is well known from literature, allowing for a correction of the beam energy in the analysis. Both approaches, however, are intrinsically limited in achievable accuracy.
Alternatively, beam-energy-independent CLS measurements can be realized by performing spectroscopy in collinear and anticollinear geometry. This has been successfully demonstrated at on-line facilities \cite{Geithner.2000, Krieger.2011, Novotny.2009} but is not generally applied for rare isotope beams due to the twice as long measurement time. 

In a CLS measurement, a high voltage is applied to an ion source. This potential defines the beam energy and can be directly measured using a voltage divider. The obtained accuracy depends on uncertainties of the divider ratio (typically $10^{-4}$ relative accuracy), contact voltages, potential gradients and field penetrations in the ion source, leading to an uncertainty of approximately $\pm\, 3-5$ eV.
The achievable accuracy of the beam-energy determination based on an isotope-shift measurement is also limited since its sensitivity is relatively low. The beam-energy dependence is chiefly caused by the relative mass difference between the isotopes. Furthermore, it depends on the transition frequency and the beam energy itself and typically achieves values between $0.1-0.5$\,MHz/eV, e.g., 0.25\,MHz/eV for $^{58,60}$Ni in the case of the $3d^9 4s\;^3\mathrm{D}_3 \rightarrow 3d^9 4p\;^3\mathrm{P}_2$ transition at 353\,nm and a beam energy of 30\,keV.

Contrarily, the rest-fame transition frequency is approximately two orders of magnitude more sensitive to an energy change (15\,MHz/eV for $^{58}$Ni) than the isotope shift, and hence, is the preferable reference for the energy determination.
For this reason, CLS-based high-voltage-metrology measurements have been proposed \cite{Poulsen.1982} to accurately measure the acceleration potential, and were realized with a relative accuracy of up to a few ppm for well-known transitions \cite{Poulsen.1988, Goette.2004, Kraemer.2018}.
In general, however, rest-frame transition frequencies are known to a few 100\,MHz in comparison to a required uncertainty of a few MHz for the determination of the beam energy at the 1-eV level, which makes this approach practically inapplicable.

In this paper we introduce an all-optical approach to determine the kinetic beam energy, which combines the advantages of the collinear-anticollinear approach and the high sensitivity of the rest-frame transition frequency on the beam energy. It does not require special equipment like a frequency comb or a precision high-voltage divider, precise literature values nor a longer measurement time. Proof-of-principle experiments were preformed on a 30-keV Ni beam reaching $10^{-5}$ relative accuracy, which corresponds to an improvement of at least one order of magnitude compared to the conventional approaches based on a high-voltage divider or an isotope-shift measurement.
In particular, collinear fluorescence and resonance-ionization spectroscopy experiments at on-line facilities can strongly benefit from the presented method.
Due to the low production rates of exotic isotopes, beam-energy independent measurements via collinear and anticollinear spectroscopy directly on these isotopes are generally not feasible due to the limited measurement time, in contrast to off-line facilities where tremendous accuracy has been demonstrated with that technique \cite{Imgram.2019, Mueller.2020}.
The procedure described here, however, can be realized with an off-line beam in preparation of the on-line experiment and then allows for \textit{in situ} beam-energy measurements during the on-line runs. Since the beam-energy uncertainty is the dominant contribution as descriptively illustrated in \cite{Heylen.2021, Lochmann.2014}, this approach can lift on-line CLS measurements on a new level of accuracy.

\section{Setup}
\label{sec:Setup}
\begin{figure}
	\centering
		\includegraphics[width=0.4800\textwidth]{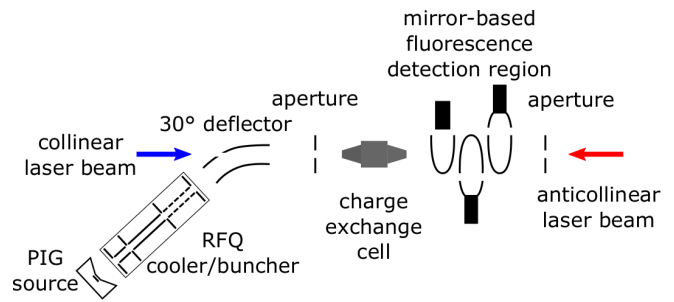}
	\caption{Schematic of the BECOLA beamline. The ion beam is produced in a Penning-ionization-gauge (PIG) source. In the radio-frequency-quadrupole trap (RFQ) the beam is cooled and can be extracted continuously or as bunches. Laser and ion beams are superimposed and aligned through two 3-mm apertures in 2.1\,m distance. Fluorescence light is collected by three mirror-based detection units. Further ion optics for beam deflection and collimation are not shown.
	}
	\label{fig:setup}
\end{figure}

A detailed description of the BEam COoling and LAser spectroscopy (BECOLA) facility at the National Superconducting Cyclotron Laboratory (NSCL) at Michigan State University can be found in \cite{Minamisono.2013, Rossi.2014}. The parts that are essential for the proposed approach are briefly presented in this section and visualized in the schematic overview shown in Fig.\,\ref{fig:setup}.
Ion beams at an energy of 30\,keV are available from either the Coupled Cyclotron Facility at NSCL (radioactive isotope beams) or from an off-line Penning-ionization-gauge (PIG) ion source \cite{Ryder.2015} at BECOLA, which is a discharge plasma sputtering source and produces stable isotope beams of predominantly singly-charged ions. The ion beams are first transported to the helium-buffer-gas-filled radio-frequency quadrupole ion trap (RFQ) \cite{Barquest.2017}, where the beams are cooled and can be accumulated and extracted as a compressed ion bunch. The bunched beam is required to perform time-resolved resonant fluorescence measurements to suppress the constant laser-induced background \cite{Nieminen.2002, Campbell.2002}. 
The laser beam is introduced in the 30$^\circ$ bender and overlaps with the ion/atom beam over the following 5-m long straight section. The beamline includes several ion optics to ensure a good alignment between laser light and ion beam, which can be checked by placing two 3-mm-diameter apertures into the beamline at a distance of 2.1\,m. Between the apertures a Na-loaded charge-exchange cell (CEC) \cite{Klose.2012} and three mirror-system-based fluorescence-detection units (FDU) \cite{Minamisono.2013, Maass.2020} are installed.
The CEC was heated to 410\,$^\circ$C to create a Na vapor leading to a 50\,\% neutralization efficiency of the incoming singly-charged ion beam through electron donation from the Na vapor. The FDUs collect a large fraction of the fluorescence light and guide it to photo-multiplier tubes that count single photon events with a time resolution of up to 16\,ns \cite{Rossi.2014}. 

The CEC is floated from the ground potential and a scanning potential is applied to perform Doppler tuning. Instead of scanning the laser frequency across the resonance, the beam velocity is adjusted by applying a small voltage $U_\mathrm{scan}$ with a full scanning width of $40$\,V that alters the beam energy ($E_\mathrm{kin} = E_\mathrm{kin,0}+eU_\mathrm{scan}$) and leads to different Doppler shifts. When the Doppler-shifted transition frequency matches the laser frequency, the atoms are resonantly excited and emit photons, which are collected by the FDUs and counted with attached photomultiplier tubes. Compared to scanning the laser frequency, Doppler-tuning enables a faster and more precise scanning procedure, chiefly because a higher stability of the laser system is achieved when operating it at a fixed frequency.

For the presented measurements, a Ni$^+$ ion beam was generated in the PIG source from natural nickel and injected into the RFQ. The Ni$^+$ beam was extracted from the RFQ at an approximate energy of 29.85\,keV in two modes. One was the bunch mode as described above, and the other was a direct current (DC) mode just passing through the RFQ without trapping and bunching. Since Ni$^+$ ions are not accessible by laser spectroscopy due to the lack of transitions in the optical regime, the ions were neutralized by collisions with sodium vapor inside the CEC. Through this non-resonant process, various electronic states are populated including the metastable  $3d^9 4s\;^3\mathrm{D}_3$ state \cite{Ryder.2015}, from which the atoms were excited to the $3d^9 4p\;^3\mathrm{P}_2$ state with laser light at 353\,nm.

The laser light was transported via two fibers to both ends of the beamline so that the ion beam could be irradiated in collinear and anticollinear geometry. A laser power of 300\,$\upmu$W was used and the laser light had a diameter of 1\,mm at the FDUs. The primary laser was a continuous-wave Ti-Sapphire laser (Matisse TS, Sirah Lasertechnik) operated at 705\,nm and pumped by a frequency-doubled Nd-YAG laser (Millennia eV, Spectra Physics). The Ti-Sapphire laser's short-term stabilization was realized by the side-of-fringe locking to a reference cavity. For long-term stabilization, the cavity length was regulated to a wavelength-meter reading (WSU30, HighFinesse), which has a specified $3\sigma$ accuracy of 30\,MHz and was calibrated every minute to a frequency-stabilized helium-neon laser (SL 03, SIOS Meßtechnik). 
The 705-nm light was sent to a cavity-based frequency doubler (Wavetrain, Spectra Physics) creating the 353-nm light that was coupled into the optical fibers and transported to the CLS beamline.

\section{Method}
In collinear (c) or anticollinear (a) laser spectroscopy measurements the resonant laser frequencies $\nu_\mathrm{c/a}$ are correlated with the beam energy $E_\mathrm{kin}$ due to the Doppler effect 
\begin{equation}
\label{eq:DopplerFormula}
\nu_\mathrm{c/a} = \nu_0 \gamma (1 \pm \beta) \approx \nu_0 \left(1 \pm \sqrt{2E_\mathrm{kin}/mc^2}\right)
\end{equation}
where $\nu_0$ is the rest-frame transition frequency, $\beta$ is the beam velocity relative to the speed of light $c$,  $\gamma=1/\sqrt{1-\beta^2}$ is the time dilation factor, and $m$ the mass of the atom. This correlation becomes more intuitive in the non-relativistic approximation. The sensitivity of the transition frequency on the beam energy is
\begin{equation}
\label{eq:diffDoppler}
\begin{split}
\frac{\partial \nu_\mathrm{c/a}}{\partial E_\mathrm{kin}} &= \frac{2 \nu_0}{mc^2} \frac{\nu_\mathrm{c/a}^2} {\nu_\mathrm{c/a}^2-\nu_0^2}  \approx \frac{\nu_0}{\sqrt{2eUmc^2}} \\
&\approx 5-30\,\mathrm{MHz/eV}
\end{split}
\end{equation}
with the total acceleration potential $U$ and the electric charge $e$. For typical experimental conditions ($10-60$\,keV, visible optical transition, medium mass atoms), a 1-eV change leads to a Doppler shift of the resonance frequency of approximately $5-30$\,MHz.  This is of the order of the natural linewidth and enables a precise determination of the beam energy. Contrarily, in the case of the isotope shift $\delta\nu^{A,A'}=\nu^A-\nu^{A'}$, the sensitivity on the beam energy $\partial\delta\nu^{A,A'}/\partial E_\mathrm{kin}$ is approximately two orders of magnitude smaller than $\partial\nu^{A}/\partial E_\mathrm{kin}$ and mainly originates from the mass difference of both isotopes.

Our approach to determine the beam energy is to make use of the rest-frame transition frequency $\nu_0$ that was separately obtained from the collinear and anticollinear laser frequencies at resonance $\nu_\mathrm{c}$ and $\nu_\mathrm{a}$, respectively. The multiplication of $\nu_\mathrm{c}$ and $\nu_\mathrm{a}$ from Eq.\ref{eq:DopplerFormula} yields the velocity- or beam-energy-independent rest-frame transition frequency 
\begin{equation}
\nu_\textnormal{c} \cdot \nu_\textnormal{a} = \nu_0^2 \gamma^2 (1+\beta) (1-\beta) = \nu_0^2 ~.
\label{Eq:ColAcol}
\end{equation}

Once the rest-frame transition frequency is determined, it can be used to extract the beam energy in combination with any collinear or anticollinear measurement performed at later times
\begin{equation}
\label{eq:Ekin}
E_\mathrm{kin} = \frac{mc^2}{2} \frac{(\nu_0-\nu_\mathrm{c/a})^2}{\nu_0 \nu_\mathrm{c/a}} ~.
\end{equation}
For example, in a typical one week long experiment on exotic isotopes, reference spectra from a stable isotope are frequently measured to determine the isotope shift. The resonance frequency of the stable isotope can then be combined with the pre-determined rest-frame transition frequency to deduce and track the drift of the beam energy.


In the present measurement, Doppler tuning was applied and the beam velocity was varied to scan across the resonance with fixed laser frequencies separately for collinear and anticollinear measurements. A small scanning potential (40\,V) was applied to the CEC to vary the otherwise constant beam energy of approximately 29.85\,keV. The scanning voltage was measured using a precision voltage divider with a relative accuracy of $6\cdot 10^{-5}$, which is negligible compared to the total beam-energy uncertainty. 
Although the collinear and anticollinear laser frequencies were chosen to be in resonance at the same scanning beam energy, a small energy difference remained. To compensate this energy difference, Eq.\,\ref{eq:Ekin} is modified, correcting one of the resonance frequencies to account for the differential Doppler shift derived in Eq.\,\ref{Eq:ColAcol}. The corrected rest-frame transition frequency is now given by

\begin{equation}
\nu_0 = \sqrt{ \left( \nu_\textnormal{c} - \frac{\partial \nu_\textnormal{c}}{\partial E_\mathrm{kin}} \cdot e \cdot \Delta U_\textnormal{scan} \right) \cdot \nu_\textnormal{a} } 
\label{Eq:ColAcol2}
\end{equation}
where $\Delta U_\mathrm{scan}$ is the difference of the scanning energies at resonance, which is typically less than 5\,eV. In the analysis, $\nu_0$ was determined iteratively starting with inserting the literature value \cite{Litzen.1993} in Eq.\,\ref{eq:diffDoppler}. If no precise literature value is available, the beam energy estimated from the set voltage can be chosen as initial guess. Sufficient convergence was already achieved by applying Eq.\,\ref{Eq:ColAcol2} to deduce a first value for $\nu_0$, inserting it in Eq.\,\ref{eq:diffDoppler} and extracting the final $\nu_0$ from Eq.\,\ref{Eq:ColAcol2}.

\begin{figure*}
	\centering
		\includegraphics[width=0.9800\textwidth]{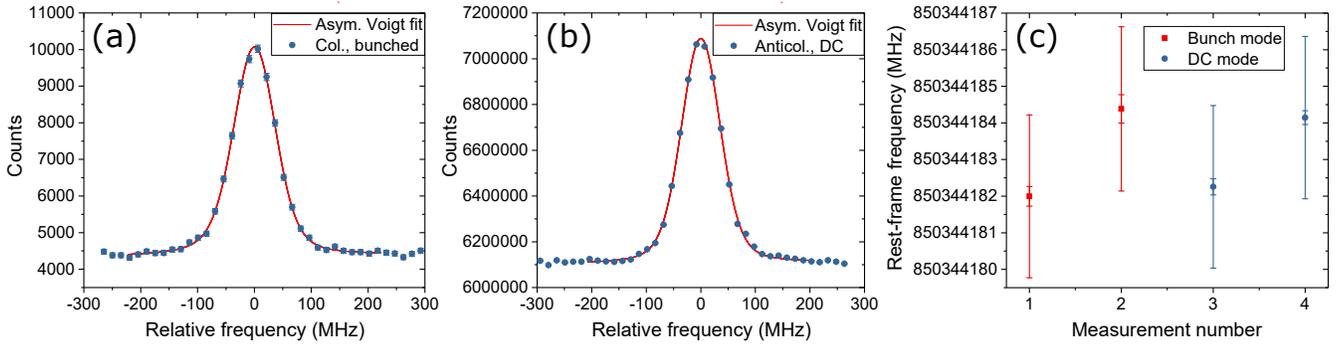}
	\caption{Typical resonance spectra of $^{60}$Ni measured in bunched mode (a) or DC mode (b). The abscissa is relative to the deduced rest-frame transition frequency of 850\,344\,183.2\,MHz.
	In the bunched mode the background rate is strongly suppressed but also signal strength is reduced compared to the DC mode since the ion beam current is limited by the capacity of the RFQ ion trap. Spectra taken in collinear or anticollinear geometry yielded a similar quality.\newline
	(c) Combining a collinear and an anticollinear measurement, the rest-frame frequency was determined. The inner error bars correspond to the fit uncertainty while the outer error bars include possible voltage drifts between the measurements, which were also considered as statistical contribution.
	}
	\label{fig:spectra}
\end{figure*}

\section{Rest-frame transition frequency determination}
\label{sec:ColAcolUnc}
The rest-frame frequencies $\nu_0$ of the $3d^9 4s\;^3\mathrm{D}_3 \rightarrow 3d^9 4p\;^3\mathrm{P}_2$ transitions in the stable $^{58}$Ni and $^{60}$Ni were determined from four collinear and anticollinear CLS measurements of each isotope. Thereof, two DC mode and two bunch mode measurements were performed. In Fig.\,\ref{fig:spectra} (a) and (b) typical spectra are depicted.
Each measurement of $\nu_0$ agreed within their statistical uncertainty as shown in Fig.\,\ref{fig:spectra} (c) for $^{60}$Ni. The averaged values are summarized in Tab.\,\ref{tab:RestFrameFreq} and are in excellent agreement with measurements from hollow cathode discharges \cite{Litzen.1993}.
The largest contribution to the present 20-MHz uncertainty originates from the frequency measurement with the wavelength meter, which will mainly cancel for the beam-energy determination. Hence, also smaller contributions are discussed in detail.

The statistically-acting uncertainties given in the first parentheses in Tab.\,\ref{tab:RestFrameFreq} consist of:
\begin{itemize}
\item \textit{Fit uncertainty:} $\leq 0.6 / \sqrt{2}$\,MHz: 
For each measurement, the photon counts of all three FDU were summed before fitting. By combining corresponding collinear and anticollinear measurements, the rest-frame frequency was extracted. The total fit uncertainty is calculated through Gaussian error propagation.
\item \textit{Voltage drifts:} 2.2\,MHz: The relative drift between the collinear and the anticollinear measurement ($\leq 1$\,h to adjust laser and frequency doubler) of the 30-kV acceleration voltage was monitored with a high-voltage divider to be below 0.3\,V. By performing the measurements in alternating order (collinear-anticollinear-anticollinear-collinear), the impact of linear drifts of the acceleration voltage (mainly temperature drifts) could be compensated and thus the uncertainty due to residual voltage fluctuations was treated statistically.  
\end{itemize}
Both contributions were added in quadrature for each of the four measurements before taking the weighted mean and applying Gaussian error propagation to calculate the statistical uncertainty. This uncertainty matches well with the standard deviation of the mean, which validates the applied procedure.

The considered systematic uncertainties were:

\begin{itemize}
\item \textit{Frequency measurement:} 20\,MHz / 1.4\,MHz: The WSU30 wavelength meter used in the present study has a 1$\sigma$ uncertainty of 10\,MHz, resulting in 20\,MHz after frequency doubling. In \cite{Verlinde.2020, Koenig.2020} this uncertainty has been investigated in more detail and it was found that it can be separated into two parts. The specified uncertainty is caused by a frequency offset that is constant over time for measurements at the same wavelength if regularly calibrated with the same reference laser. On top of this offset only relatively small variations have been observed. A 3-MHz amplitude of these local variations is quoted for a similar wavelength meter in \cite{Koenig.2020}, which is interpreted as a 1$\sigma$ uncertainty of 1\,MHz.

A constant offset $\delta \nu$ in a collinear-anticollinear frequency measurement leads to an identical shift of the rest-frame transition frequency
\begin{equation}
\label{eq:colAcol-FreqOffset}
\begin{split}
(\nu_\mathrm{c}+ \delta \nu) (\nu_\mathrm{a}+ \delta \nu) &= \nu_\mathrm{c}\nu_\mathrm{a} + \delta \nu (\nu_\mathrm{c}+\nu_\mathrm{a})+ \delta \nu^2 \\
&\approx (\nu_0 + \delta \nu)^2 
\end{split}
\end{equation}
with an approximation of $\nu_\mathrm{c}+\nu_\mathrm{a}\approx 2 \nu_0$. 
In the later discussion of the beam-energy determination the offset contribution cancels, transferring the frequency-measurement uncertainty to be caused by the local variations of the wavelength meter. They affect collinear and anticollinear measurements independently and hence, the uncertainty was determined by Gaussian error propagation 
\begin{equation}
\Delta \nu_\mathrm{0,var}= \frac{1}{\sqrt{2}} \Delta \nu_\mathrm{WM- var} \cdot 2 = 1.4\,\mathrm{MHz}~.
\end{equation}
To account for the frequency doubling, it was multiplied by a factor of two.

\begin{table}
\centering
\caption{Rest-frame transition frequencies of the $3d^9 4s\;^3\mathrm{D}_3 \rightarrow 3d^9 4p\;^3\mathrm{P}_2$ transition in the naturally most abundant neutral nickel isotopes $^{58,60}$Ni. The transition has been measured before from hollow cathode discharges but only the average over all stable isotopes is given \cite{Litzen.1993}. Isotope-separated values have been calculated according to the natural abundance and the isotope shift from \cite{Steudel.1980}, which agree closely with our results when interpreting the uncertainty in \cite{Litzen.1993} of a few mK to be about 100\,MHz. Our statistical uncertainty is given in the first parentheses while the systematic uncertainty is listed in the second parentheses.}
\begin{tabularx}{0.41\textwidth}{c c c}
\addlinespace[.6em]
\hline
\addlinespace[.2em]
\multicolumn{1}{c}{Isotope}	 & \multicolumn{1}{c}{This work} & \multicolumn{1}{c}{Literature \cite{Litzen.1993}}    \\ 
\multicolumn{1}{c}{}			 &  \multicolumn{1}{c}{MHz}               &  \multicolumn{1}{c}{MHz}                      \\ 
\addlinespace[.2em]
\hline
\addlinespace[.2em]
$^{58}$Ni                           & 850\,343\,677.6\,(1.2)\,(20.0)~~                   & 850\,343\,600\,(100)           \\
$^{60}$Ni                           & 850\,344\,183.2\,(1.1)\,(20.0)~~                 & 850\,344\,110\,(100)           \\
\addlinespace[.2em]
\hline
\end{tabularx}
\label{tab:RestFrameFreq}
\end{table}

\item \textit{Line shape:} 1\,MHz: The resonance line shape becomes asymmetric since the atoms experience an energy loss in inelastic collisions with sodium atoms in the CEC. Non-matching fit functions can lead to shifts of the extracted resonance centroid frequencies, which however, appear in opposite directions in the collinear and anticollinear measurements and cancel in the extraction of $\nu_0$. Data fitting was done by using a symmetric Voigt function, a Voigt with an additional satellite Voigt, and a Voigt with an exponential function \cite{Stancik.2008}. The observed discrepancies of $\nu_0$ are below 1\,MHz and can still have statistical origin but conservatively the largest deviation between the different fit models was considered.
\item \textit{Beam alignment:} 0.8\,MHz: The laser light paths were checked in 5\,m distance at the entrance and exit of the CLS windows. The misalignment between the collinear and anticollinear laser light was estimated to be smaller 1\,mrad. The laser light and ion beam alignment was checked with two 3-mm apertures in 2.1\,m distance leading to a maximal angular deviation of 2\,mrad. Including the angular dependence, Eq.\,\ref{Eq:ColAcol} yields 
\begin{equation}
\nu_0'^2= \nu_\mathrm{a}\nu_\mathrm{c} \gamma^2 \,(1 + \beta \cos\alpha_\mathrm{a})\,(1 - \beta \cos\alpha_\mathrm{c}) 
\end{equation}
with $\alpha_\mathrm{c}$ and $\alpha_\mathrm{a}$ being the angles between atomic beam and collinear or anticollinear laser light, respectively. Calculating the maximum frequency deviation for the available parameter space that is limited by $\alpha_\mathrm{c},\,\alpha_\mathrm{a}<2$\,mrad and $|\alpha_\mathrm{c}-\alpha_\mathrm{a}|<1$\,mrad, the largest deviation is 0.8\,MHz while the mean deviation over the whole parameter space would be 0.25\,MHz. 
\item \textit{Other:} 0\,MHz: At the current level of precision, further uncertainty contributions are neglected. \newline
\textit{Bunch structure:} The rest-frame frequencies obtained from measurements in bunch and DC mode agree well within their fit uncertainties, and hence do not indicate any systematic discrepancy.\newline
\textit{Scan voltage:} All measurements were performed at a similar scanning potential and hence, do not have significant contributions due to deflecting or focusing the beam, due to the voltage measurement, nor due to the linear approximation in Eq.\,\ref{Eq:ColAcol2}.\newline
\textit{Beam overlap:} If the laser beams differ in position or diameter, they can interact with different parts of the atomic beam. Due to the beam cooling in the RFQ and the resulting homogeneous atomic beam, the estimated impact is negligible. \newline
\textit{Photon recoil:} With each laser-atom interaction, a directed moment is transferred to the atom while the emittance of fluorescence light is undirected. This leads to an acceleration of the atoms if the atomic and the laser beam are parallel and to a deceleration if both beams have opposite direction, which contradicts to the requirement of a constant beam energy of Eq.\,\ref{Eq:ColAcol}. Comparing the ratios of the different detection units for both cases did not show any systematic trend at our current resolution.\newline
\textit{Optical population transfer:} The applied transition is not a two-level system and 10\,\% of the excited atoms will decay into a dark state. Hence, multiple interactions in front of the optical detection regions will depopulate the ground state, especially for the resonance condition. We assume that we cover this effect within the line shape contribution.
\end{itemize}
The total 20-MHz uncertainty of the rest-frame frequency determination is dominated by the uncertainty of the laser-frequency measurement with the wavelength meter. The smaller contributions will become significant for the beam-energy determination, where the frequency-measurement uncertainly can be mostly eliminated.

\section{Beam-energy determination}
\label{sec:CalibrationUnc}
Again, the specified wavelength meter uncertainty is separated into a constant offset and local variations.
Including a frequency offset $\delta \nu$, Eq.\,\ref{eq:Ekin} yields
\begin{equation}
\label{eq:EkinOffsetUnc}
\begin{split}
E_\mathrm{kin} &= \frac{mc^2}{2} \frac{((\nu_0+\delta \nu)-(\nu_\mathrm{c/a}+\delta \nu))^2}{(\nu_0+\delta \nu) (\nu_\mathrm{c/a}+\delta \nu)} \\
&= \frac{mc^2}{2} \frac{(\nu_0-\nu_\mathrm{c/a})^2}{\nu_0 \nu_\mathrm{c/a}+\delta \nu(\nu_0 + \nu_\mathrm{c/a}) + \delta \nu^2}
\end{split}
\end{equation}
where the contribution from $\delta \nu$ mostly cancels and falls below the $10^{-7}$ level, and hence, is not the dominant factor for the precise determination of the beam energy.
The elimination of $\delta \nu$, is based on the use of the same wavelength meter for the determination of $\nu_0$ and the independent measurements of $\nu_\mathrm{a/c}$. If a literature value is to be used, the wavelength-meter-offset contribution does not cancel, leading to significantly larger uncertainties in the energy determination.

The uncertainty in the frequency difference $(\nu_0-\nu_\mathrm{c/a})$ in the numerator of Eq.\,\ref{eq:Ekin} becomes now the dominant contribution for the beam-energy determination.
Since the uncertainties of $\nu_0$ have been discussed in detail in the previous section, we will now focus on $\nu_\mathrm{c/a}$:
\begin{itemize}
\item \textit{Fit uncertainty:} $\leq 0.6$\,MHz: Uncertainty given by the fit of the function that is best suited for describing the data. In our case, this was the asymmetric Voigt function that contains an exponential function to describe the slightly asymmetric resonance line shape.
\item \textit{Line shape:} 1.5\,MHz: Comparing the resonance frequencies obtained with the three fit functions discussed above, deviations $<1.5$\,MHz between asymmetric fit functions and $<5$\,MHz between symmetric and asymmetric fit functions were observed. Since the measured spectra were clearly asymmetric, the deviation between the asymmetric fit functions was considered.
\item \textit{Local wavelength-meter variations:} 2\,MHz: As discussed in section\,\ref{sec:ColAcolUnc}, a 1-MHz uncertainty of the wavelength-meter reading was considered. To account for frequency doubling, this value was multiplied by a factor of two.
\item \textit{He:Ne drift:} 2\,MHz: Day-to-day drifts of the Helium-Neon laser frequency used for calibrating the wavelength meter may vary.
\item \textit{Beam alignment:} 0\,MHz: This method is directly extracting the velocity component along the laser direction. Therefore, no misalignment has to be considered.
\end{itemize}
As for the uncertainties of $\nu_0$, all contributions discussed in section\,\ref{sec:ColAcolUnc} except the wavelength-meter offset were included. 
Adding these contributions in quadrature, a total uncertainty of $\Delta (\nu_0-\nu_{c/a}) =4.0$\,MHz was obtained, which corresponds to a kinetic-energy uncertainty of 0.27\,eV and hence, enabled a $9 \cdot 10^{-6}$ measurement of the Ni beam energy at 29.85\,keV.

\begin{figure}
	\centering
		\includegraphics[width=0.4800\textwidth]{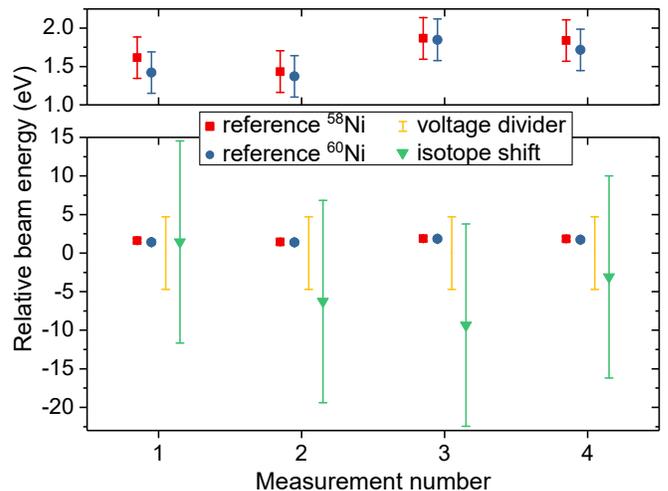}
	\caption{Kinetic beam energy deduced by three different approaches relative to the set point of the power supply employed for beam acceleration (29850\,V). The uncertainty of the present method is more than one order of magnitude smaller, and hence, the results for measurements based on transitions in $^{58}$Ni and $^{60}$Ni are also shown in higher resolution in the upper part of the figure. The results based on the isotope shift rely on the same experimental data but yield a higher uncertainty due to the lower sensitivity of this approach. Due to an overdue calibration of the high-voltage divider, only the nominal uncertainty of this approach is plotted. 
	}
	\label{fig:Results}
\end{figure}

\section{Discussion}
In Fig.\,\ref{fig:Results}, our approach is compared to the conventional methods.
For demonstration purposes, four $^{58}$Ni and four $^{60}$Ni measurements were performed in alternating order in collinear geometry.
As depicted in the upper part of the Fig.\,\ref{fig:Results}, the present approach shows consistent results for both isotopes. 
The deviation between both isotopes varies between 0.02\,eV and 0.2\,eV, which is caused by statistical uncertainties and chiefly voltage drifts between the measurements. After measurement set (2) was a 1-h time break explaining the larger step.

In the lower part of the Fig.\,\ref{fig:Results}, all methods are compared yielding an excellent agreement. However, the uncertainties of the conventional methods are significantly larger than those of the present approach.

To demonstrate the isotope-shift-based approach, the isotope shift between $^{58,60}$Ni of the same set of measurements was evaluated and the beam energy was adjusted in the analysis until the isotope shift matched the literature value of $\delta \nu (^{60,58}\mathrm{Ni})=507.8\,(0.9)$\,MHz \cite{Steudel.1980}. The beam energies, for which an agreement between the measured isotope shifts and the literature value was achieved, are plotted in Fig.\,\ref{fig:Results}.
Adding the fit uncertainty ($\leq 0.6$\,MHz $\cdot \sqrt{2}$), the contributions related to the local wavelength-meter variations (2\,MHz $\cdot \sqrt{2}$) and the line shape (1\,MHz) in quadrature to the uncertainty of the literature value, yields a combined uncertainty of $\Delta \delta\nu^{AA'}=3.2$\,MHz corresponding to $\Delta E_\mathrm{kin}=13$\,eV due to the much lower sensitivity of the isotope shift on the beam energy.
Furthermore, this method critically depends on a stable beam energy between the measurements of both isotopes, which explains the scatter in Fig.\,\ref{fig:Results}. The beam-energy differences observed between both isotopes with the transition-frequency-based approach were in the range of 0.18\,eV, which seems to be minor but this is amplified by $\partial \nu^A / \partial E_\mathrm{kin} \cdot (\partial \delta\nu^{AA'} / \partial E_\mathrm{kin})^{-1}\approx 60$ and causes fluctuations of 11\,eV in the isotope-shift-based approach in the case of $^{58,60}$Ni.

Using a high-voltage divider to measure the acceleration potential that defines the beam energy, is limited by the uncertainty of the divider ratio (Ohmlabs HVS-100, originally specified relative accuracy $8\cdot 10^{-5}$) and of the voltmeter (Keysight 34465A, $6\cdot 10^{-5}$). The trapping potential well in the RFQ had a nominal depth of -4\,V and a release potential of -15\,V was applied. Since the field penetration during the extraction of the trapped ions is not exactly known, a 3-V uncertainty was considered. In addition, a 2-V uncertainty was included to regard contact and thermal potentials at the RFQ and the hot CEC, leading to a total uncertainty of 4.7\,V. The calibration of the available high-voltage divider is long outdated and a significant change of the divider ratio has been observed by comparing it to the set voltage. Hence, the absolute voltage values could not be accurately evaluated with this device while relative values were still valid and used to estimate the voltage fluctuation over the measurement period. Therefore, only the size of the uncertainty based on a valid calibration is shown in Fig.\,\ref{fig:Results} and the center value is defined by the set voltage.

\section{Conclusion}
An approach to determine the kinetic energy of an atom, ion or molecule beam for collinear laser spectroscopy measurements was demonstrated using a 30-keV Ni beam. The rest-frame transition frequencies of $^{58,60}$Ni were determined by collinear and anticollinear laser spectroscopy and used as a reference to deduce the beam energy. This method has several advantages compared to conventional approaches:
\begin{itemize}
    \item \textit{High accuracy} at the $10^{-5}$ level, corresponding to an increase by more than one order of magnitude.
    \item \textit{No special equipment} like a precision voltage divider or a frequency comb are required.
    \item \textit{No assumptions} on energy shifts due to field penetrations, or due to contact and thermal potentials.
    \item \textit{No dependence} on literature values.
    \item \textit{No additional on-line measurement time} as required for the beam-energy-independent measurements, e.g., in \cite{Geithner.2008, Noertershaueser.2009, Krieger.2012}.
\end{itemize}
The application of the presented method to determine the kinetic beam energy will significantly improve the accuracy of collinear fluorescence and resonance-ionization-spectroscopy measurements on rare isotope beams by transforming the formerly largest systematic uncertainty into a minor contribution.

\section{Acknowledgements}
We acknowledge support by the National Science Foundation grant No. PHY-15-65546.

\bibliographystyle{aipnum4-1}
\bibliography{literature}
\end{document}